\documentclass[11pt,twoside]{article}
\usepackage{asp2010}

\def\deg{^\circ}             %for angular measure in degrees
\def\msun{{\rm\,M_\odot}}

\resetcounters

\begin{document}

\title{Formation Models of the Galactic Bulge}
\author{Ortwin Gerhard$^1$
\affil{$^1$Max Planck Institute for extraterrestrial Physics,  
       PO Box 1312, Giessenbachstr., 85741 Garching, Germany}}

\begin{abstract}
  The Galactic bulge is now considered to be the inner
  three-dimensional part of the Milky Way's bar. It has a peanut shape
  and is characterized by cylindrical rotation. In N-body simulations,
  box/peanut bulges arise from disks through bar and buckling
  instabilities. Models of this kind explain much of the structure and
  kinematics of the Galactic bulge and, in principle, also its
  vertical metallicity gradient. Cosmological disk galaxy formation
  models with high resolution and improved feedback models are now
  able to generate late-type disk galaxies with disk-like or barred
  bulges. These bulges often contain an early collapse stellar
  population and a population driven by later disk instabilities. Due
  to the inside-out disk formation, these bulges can be predominantly
  old, similar to the Milky Way bulge.

\end{abstract}

\section{Introduction: the barred Galactic bulge}

The Galactic bulge, subject of many studies here at CTIO observatory,
is now considered to be the inner three-dimensional part of the Milky
Way's bar. Its triaxial shape was first established by the COBE NIR
photometry \citep{Blitz+Spergel91, Dwek+95, Binney+97,
  Bissantz+Gerhard02}.  Star counts confirmed this but also pointed to
a larger, in-plane bar, the so-called 'long bar' \citep{Stanek+94,
  Lopez-Corredoira+05, Skrutskie+06, Benjamin+05,
  Cabrera-Lavers+07b}. Comparing the observed HI and CO lv-diagrams
with hydrodynamic models showed that many of the observed features
could be naturally, sometimes even quantitatively interpreted with gas
flow models in barred potentials, modeled on the COBE data
\citep{Englmaier+Gerhard99, Fux99, Bissantz+03,
  Rodriguez-Fernandez+Combes08}.

In recent years, Galactic bulge studies have been revitalized by new
large photometric and spectroscopic surveys. With these new data, the
density and metallicity structure of the Milky Way's bulge and its
kinematics and dynamics have been mapped in much greater detail than
possible before. Comparison of this information with models of bulge
formation is promising to give us a better understanding of the origin
of the Galactic bulge.

\section{Box/peanut bulge models from disk instability}

N-body simulations following the evolution of isolated disk galaxies
have shown that stellar disks are often unstable to bar formation
\citep{Sellwood81, Athanassoula02, Debattista+06}.  Subsequently, the
bar may quickly go through a second, so-called buckling instability,
resulting in the formation of an inner boxy bulge
\citep{Combes+Sanders81, Combes+90, Raha+91}.  The bar may then evolve
by losing angular momentum to the dark matter halo, grow in size, and
eventually go through a second buckling instability. This results in a
a strongly peanut-shaped bulge \citep{Martinez-Valpuesta+06} much like
those seen in barred galaxies \citep{Athanassoula05}, which is
supported by three-dimensional 2:1:2 resonant orbit families that
support a characteristic X-shape \citep{Pfenniger+Friedli91}.

Recent star count and stellar kinematics data have shown that the
predictions of these idealized models characterize the properties of
the Milky Way bulge surprisingly well. The apparent magnitude plot for
red clump stars in high-latitude pencil beams shows a clearly bimodal
distance distribution, termed the 'split red clump'. This was first
seen in 2MASS data \citep{McWilliam+Zoccali10, Saito+11} but
has also been confirmed with OGLE data \citep{Nataf+10} and the
ARGOS spectroscopic bulge sample \citep{Ness+12}. The split red
clump is a signature of the X-shaped stellar orbits in the bulge.
Recent deconvolution of the VVV DR1 star count data has shown that the
Galactic bulge has the shape of a highly elongated bar with a strong
peanut shape \citep{Wegg+Gerhard13}. The measured angle of the
barred bulge to the line of sight is $(27 \pm 2)\deg$. Along the bar
axes the density falls off roughly exponentially, with axis ratios (10
: 6.3 : 2.6) and exponential scale-lengths (0.70 : 0.44 : 0.18) kpc,
but along the major axis the profile becomes shallower beyond 1.5 kpc.

N-body models with box/peanut bulges grown from unstable disks have
been used to interpret not only the structure, but also the stellar
kinematics and metallicity distribution of the Milky Way bulge.
\citet{Gerhard+Martinez-Valpuesta12} showed that the flattening in the
longitude profiles seen for $\vert l \vert < 4\deg$ in low-latitude
star counts \citep{Nishiyama+05, Gonzalez+11b} is naturally reproduced
by the density distribution in a simulated box/peanut
bulge. \citet{Li+Shen12} showed that their N-body model shows a
similar split red clump as the Galactic bulge when viewed at an
appropriate orientation. N-body bulges also explain the
near-cylindrical rotation and the velocity dispersion profiles seen in
the BRAVA \citep{Kunder+12} and Argos \citep{Ness+13a} spectroscopic
surveys \citep{Shen+10, Ness+13b}.

The match of the models to the kinematic data is very good and
\citet{Shen+10} concluded that it would be worsened even by a fairly
small additional classical bulge. Such a classical bulge could have
arisen from an early merger episode that preceded the growth and
subsequent instability of the early disk. \citet{Saha+12, Saha+13}
showed that angular momentum transfer from the bar and boxy bulge can
spin up the preexisting bulge, so that it too develops cylindrical
rotation.  The radial dependence of the rotation is different between
the two bulges though; thus some more work is needed on this issue.

The strongest argument for a classical bulge in the Milky Way has been
the observation of a strong vertical metallicity gradient
\citep{Zoccali+08, Johnson+11, Gonzalez+13}. However, a recent
analysis of an N-body bulge showed that this argument is not
compelling.  If the Milky Way's bar and bulge formed rapidly from the
disk at early times, then any preexisting metallicity gradients are
preserved as vertical gradients in the final bulge, because of the
incomplete violent relaxation during the instabilities
\citep{Martinez-Valpuesta+Gerhard13}. Therefore, at present there is
no strong evidence for a classical bulge in the Milky Way. Further
chemodynamical analysis of the metallicity distribution and kinematics
will be needed to clarify whether the Milky Way has a small classical
bulge underneath its dominant peanut-shaped bulge, and if so how this
is related to the bulge metallicity components identified by
\citet{Ness+13a} in the ARGOS sample.

The connection between the Milky Way's peanut bulge and planar bar is
much less explored, primarily because this region is observationally
less accessible. \citet{Hammersley+00, Benjamin+05,
  Cabrera-Lavers+07b} found evidence from starcounts for a 'long bar'
at an angle of $\sim 45\deg$, apparently tilted relative to and
therefore distinct from the barred bulge.
\citet{Martinez-Valpuesta+Gerhard11} used an N-body model to argue
that the long bar may nonetheless be the planar part of the {\sl same}
Galactic bar that contains the peanut bulge \citep[see
also][]{Romero-Gomez+11}. In their model, the outer parts of the bar
oscillate from trailing to leading, coupling with nearby spiral arms
which rotate at lower pattern speed \citep[][]{Tagger+87}. Future
APOGEE kinematic observations in this region
\citep[e.g.][]{Nidever+12} will test the model predictions and shed
more light on the dynamical transition between bulge, two-dimensional
bar, and inner disk.

\section{Bulges in cosmological disk galaxy simulations}

Galaxies in the hierarchical universe grow continuously by accretion
of matter, and they are subjected to a variety of external
perturbations, for example, from merging satellites. Therefore,
eventually the idealized disk simulations will need to be superceded
by more realistic disk galaxy formation simulations embedded in a
cosmological setting.  It has only become possible recently to follow
realistic high-resolution, fully cosmological simulations from high
redshift to now. These models are already stimulating our
understanding of bulge formation, but they still have considerable
uncertainties in how the star formation and feedback processes are
modelled.

As a reference point, I start with a brief discussion of the
dissipative collapse model by \citet{Samland+Gerhard03}. This may be
viewed as a modern version of the dissipative collapse envisaged by
\citet{Eggen+62}. The model considered the
formation of a large disk galaxy in a $2\times 10^{12} \msun$,
spinning ($\lambda=0.05$) halo, with accretion history taken from
cosmological simulations. Star formation and feedback were modelled by
a 2-phase fluid/cloud fluid model calibrated on observed star
formation rates, and the enrichment of Fe and $\alpha$-elements was
tracked separately. Axisymmetric collapse was followed excluding all
minor mergers.

In this dissipative collapse, the galaxy formed from inside out. The
star formation history closely followed the gas accretion history.
The disk grew from initially small, thick and dynamically hot to a
large, thin and cold disk at redshift $z=0$. The final bulge consisted
of at least two stellar populations. One was formed in the early rapid
collapse phase, the second was formed later in the disk which grew bar
unstable. They differed by their [$\alpha$/Fe]-ratios; the final bulge
contained [$\alpha$/Fe]-enhanced stars for a range of [Fe/H], and also
some super-solar stars made from gas channeled inwards by the bar.
This illustrates the chemodynamical signatures which can be used to
understand the origin of the Milky Way bulge from the new data.

\citet{Obreja+13} compared the properties of bulges made in some of
the recent cosmological simulations, from \citet{Brook+12,
  Domenech-Moral+12}. Generally, these simulations employ strong
feedback to prevent overproduction of stars at early times. As in the
dissipative collapse model, the star formation rate (SFR) in these
simulations closely follows the mass accretion rate, and the formation
history can be divided into an early starburst-collapse phase and a
later phase with lower SFR driven by disk instabilities and minor
mergers.  The stellar populations show corresponding age and
metallicity distributions, with the old population being more metal
poor and $\alpha$-enhanced.  Most of the old population found in the
final bulge formed in disjoint places along filaments at high redshift
(as opposed to in the early collapsing halo in the dissipative
collapse model).

One can associate the rapid phase with a classical bulge, and the late
phase with a box/peanut or disky bulge. Such a classical bulge is
clearly different from a bulge formed by a late major merger. In all
of the five simulations analyzed by \citet{Obreja+13}, both components
were present, with different mass ratio. \citet{Guedes+13} and
\citet{Okamoto+13} followed the evolution of such 'pseudo-bulges' in
very high resolution simulations. The structure of these bulges is
characterized by a Sersic index around $n=1\pm 0.5$, possibly
depending on the feedback model. One important result from these
simulations is that, because of the inside-out formation, the inner
bulge region can form early and consist of predominantly very old
stars, whereas the disk further out has a much more uniform star formation
history. In the Eris simulation of Guedes et al, the bulge is built
from the early, low-angular momentum part of the disk. This model does
not have a final box/peanut bulge like in the Milky Way, but it
illustrates how the bulge stars could be very old as is observed in
Baade's window in the Galactic bulge.

\bibliographystyle{asp2010}

%\bibliography{MW+DiskGal_2013Sep29}
\bibliography{Gerhard.bbl}

\end{document}